\documentclass[10pt,conference]{IEEEtran}
\IEEEoverridecommandlockouts
\usepackage{cite}
\usepackage{amsmath,amssymb,amsfonts}
\usepackage{algorithmic}
\usepackage{graphicx}
\usepackage{todonotes}
\usepackage{textcomp}
\usepackage{xcolor}
\def\BibTeX{{\rm B\kern-.05em{\sc i\kern-.025em b}\kern-.08em
    T\kern-.1667em\lower.7ex\hbox{E}\kern-.125emX}}
    
\usepackage{quantikz}
\usepackage{algorithm}
\usepackage{algorithmic}
\usepackage{hyperref}
\usepackage{subcaption}
\usepackage{xcolor}

\begin{document}

\title{Quantum Approximate Optimization Algorithm with Sparsified Phase Operator 

\thanks{This work was supported in part by the U.S.\ Department of Energy (DOE), Office of Science, Office of Advanced Scientific Computing Research AIDE-QC and FAR-QC projects and by the Argonne LDRD program under contract number DE-AC02-06CH11357.}
}

\author{\IEEEauthorblockN{Xiaoyuan Liu\IEEEauthorrefmark{1}\IEEEauthorrefmark{2}, Ruslan Shaydulin \IEEEauthorrefmark{3}, Ilya Safro\IEEEauthorrefmark{2}}
\IEEEauthorblockA{\IEEEauthorrefmark{1}Fujitsu Research of America, Inc., Sunnyvale CA, USA}
\IEEEauthorblockA{\IEEEauthorrefmark{2}Department of Computer and Information Sciences, University of Delaware, Newark DE, USA}
\IEEEauthorblockA{\IEEEauthorrefmark{3}Mathematics and Computer Science Division, Argonne National Laboratory, Lemont IL, USA}
}

\maketitle

\begin{abstract}
The Quantum Approximate Optimization Algorithm (QAOA) is a promising candidate algorithm for demonstrating quantum advantage in optimization using near-term quantum computers. However, QAOA has high requirements on gate fidelity due to the need to encode the objective function in the phase separating operator, requiring a large number of gates that potentially do not match the hardware connectivity. Using the MaxCut problem as the target, we demonstrate numerically that an easier way to implement an alternative phase operator can be used in lieu of the phase operator encoding the objective function, as long as the ground state is the same. We observe that if the ground state energy is not preserved,  the approximation ratio obtained by QAOA with such phase separating operator is likely to decrease. Moreover, we show that a better alignment of the low energy subspace of the alternative operator leads to better performance. Leveraging these observations, we propose a sparsification strategy that reduces the resource requirements of QAOA. We also compare our sparsification strategy with some other classical graph sparsification methods, and demonstrate the efficacy of our approach.\\
Reproducibility: Source code and data are available at \href{https://github.com/JoeyXLiu/qaoa-sparse}{https://github.com/JoeyXLiu/qaoa-sparse} \\
\end{abstract}


\section{Introduction}
Hybrid quantum-classical algorithms such as the Quantum Approximate Optimization Algorithm (QAOA) \cite{Hogg2000,farhi2014quantum} are promising candidates to demonstrate quantum advantage in solving optimization problems on near-term quantum devices. However, such factors as the hardware connectivity, high error rates, and limited fidelity of the gates on near-term quantum devices significantly restrict the performance of such algorithms, especially when the depth of the circuit is growing. Recent experiments \cite{harrigan2021quantum} demonstrate a great potential of QAOA when the problem is tailored to the hardware. For more general instances of optimization problems that cannot be mapped to hardware-native topologies, the noise significantly impacts the performance of QAOA.

QAOA is a hybrid quantum-classical algorithm for approximately solving combinatorial optimization problems. At each call to the quantum computer, a trial state is prepared by applying a sequence of pairs of alternating quantum operators. The two alternating operators are the phase operator, which encodes the objective function of the combinatorial optimization problem, and the mixing operator. Various approaches have been studied with the aim of improving QAOA. These approaches include  improving QAOA parameters' optimization \cite{zhou2020quantum,galda2021transferability,Streif2020,shaydulin2019multistart,Shaydulin2020Symmetries,shaydulin2022parameter,Shaydulin2021symmtrain,2108.13056,khairy2019learning}, designing different mixing operators \cite{hadfield2017quantum,zhu2020adaptive,bartschi2020grover}, using different initialization strategies \cite{egger2021warm,tate2020bridging,2203.13936} and  cost functions \cite{Barkoutsos2020}, introducing error mitigation schemes~\cite{shaydulin2021error}, introducing additional parameters~\cite{herrman2021multi}, construct problem specific instance-specific ansatz~\cite{wurtz2021classically}, to mention just a few. Many of these approaches aim to achieve a faster convergence with smaller circuit depth, thus reducing the quantum resource requirements of the algorithm. 

In this work, we perform modifications to the phase operator in the QAOA circuit with the same goal of reducing resource requirements. Specifically, we investigate the impact of using a phase operator with a Hamiltonian other than the one encoding the objective function. We study QAOA in the context of MaxCut, and demonstrate that the gate count of the phase operator can be reduced by sparsifying the problem graph. We observe that the performance of QAOA does not deteriorate as long as the sparsified graph has the same optimal solution, or, equivalently, as long as the corresponding Hamiltonian has the same ground state. Furthermore, if the low energy states of Hamiltonians do not align, the QAOA performance deteriorates. We numerically investigate a number of graph sparsification techniques and demonstrate their efficacy in reducing the cost of implementing QAOA for MaxCut.

\section{Background}\label{sec:background}
\subsection{QAOA}
We first briefly review QAOA. Suppose we have a  combinatorial optimization problem that is specified by the objective function $C(x), ~x\in \{0, 1\}^n$. The goal is to find a string of bits $x$ that maximizes $C(x)$. The QAOA is a variational quantum algorithm for approximate optimization which aims to find a string of bits $x$ such that $C(x)$ is close to $C_{\max}$, the maximum of the objective function $C(x)$.

The QAOA circuit prepares the trial state by applying a sequence of alternating parameterized phase and mixing operator unitaries. The phase operator is defined as $U(\mathcal{H}, \gamma) = e^{-i\gamma\mathcal{H}}$ and the mixing operator as $U(B, \beta) = e^{-i\beta B}$. Here, $\mathcal{H}$ is the Hamiltonian operator that encodes the objective function value of the optimization problem in the computational basis state, and $B = \sum_{j=1}^n \sigma_j^x$ is defined as the operator of $\sigma^x$ acting on each qubit. 

The trial state of a $p$-level QAOA is then given by: \begin{align}\label{eq:qaoa_trial_state}
\ket{\gamma, \beta} = U(B, \beta_p)U(\mathcal{H}, \gamma_p) \cdots U(B, \beta_1)U(\mathcal{H}, \gamma_1) \ket{+}^{\otimes n},
\end{align}
where $\ket{+}^{\otimes n}$ is the uniform superposition of computational basis states. After the parameterized state preparation, we measure the trial state in the computational basis. The state preparation and measurement are iteratively repeated. At each iteration, the output is used by an outer optimization loop to find good parameters $\gamma, \beta$ to maximize the objective function
\begin{align}
\max_{\gamma, \beta} ~\langle\mathcal{H}\rangle_p = \bra{\gamma, \beta} \mathcal{H} \ket{\gamma, \beta}.
\end{align}
The outer optimization loop (also known as main driving routine) is executed on a classical machine. 

To evaluate the quality of the final quantum state $\ket{\gamma, \beta}$, we compute the approximation ratio $\rho$ defined as follows:
\begin{align}\label{eq:approximationratio}
\rho = \frac{\langle\mathcal{H}\rangle_p}{C_{\max}}.
\end{align}

\subsection{MaxCut problem}
Given an undirected, unweighted graph $G = (V, E)$, where $V$ is the set of $n$ vertices and $E$ is the set of $m$ edges. The goal of MaxCut is to partition the graph into two disjoint sets, such that the number of edges between the two sets is maximized. A corresponding Hamiltonian $\mathcal{H}$ for MaxCut is defined as: \begin{align}\label{eq:maxcut_hamiltonian}
\mathcal{H} = \frac{1}{2} \sum_{ij\in E}(I - \sigma_i^z\sigma_j^z),
\end{align}
where $I$ is the identity operator, and $\sigma_i^z$ is the Pauli-$Z$ operator acts on qubit $i$. Qubits $i$ and $j$ correspond to the respective nodes connected by edge $ij\in E$.

\subsection{QAOA circuit for MaxCut}
The trial state of QAOA on MaxCut produces the state (\ref{eq:qaoa_trial_state}) with the problem Hamiltonian given in  (\ref{eq:maxcut_hamiltonian}). There exist several different ways to implement QAOA. Typically, we assume that a given basis gate set is $\{R_x, R_y, R_z, \text{CNOT}, \text{H}\}$, where $\text{H}$ is the Hadamard gate. 

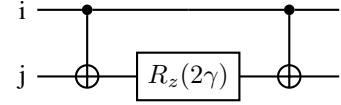
\begin{figure}[htbp!]
\centering
\begin{quantikz}
\lstick{i} & \ctrl{1} & \qw & \ctrl{1} & \qw \\
\lstick{j} & \targ{}  & \gate{R_z(2\gamma)} & \targ{} & \qw
\end{quantikz}
\caption{Circuit for edge $ij$ in the phase operator}
\label{fig:circuit}
\end{figure}

The QAOA circuit for MaxCut consists of three parts. The first part gives us the initial state that is the uniform superposition over all possible states. This is implemented by the Hadamard gates acting on each qubit. The second part is the phase operator $U(\mathcal{H}, \gamma) $. Note that \begin{align*}
U(\mathcal{H}, \gamma) &= e^{-i\gamma(\frac{1}{2} \sum_{ij\in E}(I - \sigma_i^z\sigma_j^z))} \\
&= \prod_{ij\in E}e^{-i\gamma(\frac{1}{2}(I - \sigma_i^z\sigma_j^z)))},
\end{align*}
where the operator $e^{-i\gamma(\frac{1}{2}(I - \sigma_i^z\sigma_j^z)))}$ that acts on qubit $i$ and qubit $j$ is implemented as shown in Fig.~\ref{fig:circuit}. Finally, the mixing operator $U(B, \beta)$ is implemented by having $R_x(\beta)$ gate acting on each qubit.

\subsection{Graph Sparsification}

The goal of graph sparsification algorithms is to select a representative sample of a given graph such that some properties of this graph are preserved. Generally, most sparsification problems can be expressed as follows: Given a graph $G_o$ and $\epsilon$, find a sparsified graph $G_s$ such that 
\[
(1-\epsilon)f(G_o) \leq f(G_s) \leq (1+\epsilon)f(G_o),
\]
where $f(\cdot)$ is a function that computes some property of a graph. Since many problems on graphs are computationally hard for given resources (e.g., because of the data size, time or space complexity of $f$ and limited available hardware), such sparsified graphs are often useful in practice. In other words, central to sparsification is the idea that we can expect that $f(G_s)$ will be sufficiently similar to $f(G_o)$ while the algorithm to compute $f$ will run much faster on $G_s$. We investigate if such methods can be used in our Sparse QAOA framework which, overall, constitutes a resource saving goal, similar to that in the classical sparsification. We now briefly review different graph sparsification methods we investigate in this paper. All of them are based on edge sparsification.

\paragraph*{Random Edge}
When comparing different edge sparsification algorithms, random edge selection is the most obvious approach and therefore can always be considered as an important baseline. We assign each edge a random value in $[0, 1]$, then we remove the edges based on the ranking of these values until we reach the target ratio of sparsification. (In the plots, we label this method as ``random''.)

\paragraph*{Algebraic Distance}
Algebraic distance ($\alpha$) \cite{john2017single,chen2011algebraic,ron2011relaxation} is a metric that quantifies the spectral distance between two nodes $i$ and $j$ in the relaxed homogeneous system for the graph Laplacian. Intuitively, it measures how dense is the graph region that connects $i$ and $j$. It can be used to rank the real ``strength'' of the edges, namely, $\alpha(i, j)$ is small when edge $ij$ connects nodes within the same dense region, and large when $ij$ is a shortcut connecting the distant regions in the graph. Various versions of the algebraic distance have been used a lot in multilevel and multigrid algorithms \cite{livne2012lean,safro2011multiscale}. We compute $\alpha$ for each edge and then remove the edges based on the ranking of these values until we reach the target ratio. (In the plots, we label this method as ``algebraic'', ``algebraic\_1'' means the normalization parameter is set to 1, ``algebraic\_1\_reverse'' means we sort the edge scores in ascending order.)

\paragraph*{Edge Forest Fire}
Edge forest fire \cite{hamann2016structure} is a variant of the forest fire node sampling approach \cite{leskovec2006sampling}. It is based on random walks. The vertices are burned starting from a random vertex, and may spread to the neighbors of a burning vertex. The intuition is that the edges and vertices that are visited more often during the random walk are more important in the graph. We can filter the edges based on the frequency of visits to each edge and sparsify the graph until we reach the target ratio. (In the plots, we label this method as ``fire''.)

\paragraph*{Local Degree}
Local degree \cite{hamann2016structure} is inspired by the idea of hub nodes, i.e., the vertices with locally relatively high degree. For each node, we will keep the edges that are connected to the neighbors that are having higher degree, so the edges left are considered to be the hub backbone of the graph. (In the plots, we label this method as ``degree''.)

\paragraph*{Local Similarity}
Local similarity \cite{satuluri2011local} uses the Jaccard measure to quantify the overlap between the neighborhood of vertices $u$ and $v$. This metric is used to rank the edges, and the edges with low overlap are removed until we reach the target ratio. (In the plots, we label this method as ``similarity''.)

\paragraph*{SCAN Structural Similarity}
Structural Clustering Algorithm for Networks (SCAN) \cite{xu2007scan} is an algorithm whose goal is to find clusters and outliers in a large network, the SCAN structural similarity metric of vertex $i$ and $j$ is computed by normalizing the number of common neighbors of $i$ and $j$ by the geometric mean of the two neighborhood's size. This metric is used to rank the edges, and the edges with low similarity are removed until we reach the target ratio. (In the plots, we label this method as ``scan''.)

\paragraph*{Simmelian backbones}
Simmelian backbones \cite{nick2013simmelian} aims to identify the edges that are connecting different dense subgraphs and those that are within the subgraphs. We use the Simmelian overlapping metric to rank the edges, and the edges with low overlap are removed until we reach the target ratio. (In the plots, we label this method as ``simmelian''.)

\paragraph*{Effective Resistance}
Graph sparsification by effective resistance \cite{spielman2011graph} is a method to include each edge of $G$ in the sparsified graph with probability proportional to its effective resistance. The effective resistance of an edge is the probability that the edge appears in a random spanning tree of $G$. The sparsified graph aims to preserve the cut of the graph. (In the plots, we label this method as ``effective''.)

\section{Graph Sparsification and Quantum Approximate Optimization Algorithm}
\label{sec:sparse}

In this section, we first discuss the motivation of combining graph sparsification and QAOA in Section \ref{sec:motivation} and propose Sparse QAOA and random Sparse QAOA in Section \ref{sec:sparseqaoamaxcut}. We then investigate the performance of Sparse QAOA and compare our Sparse QAOA with several other sparsification methods in Section \ref{sec:sparsification}. In Section \ref{sec:en_alignment} we analyze the relationship between the QAOA approximation ratio and the alignment of low energy subspaces. In Section \ref{sec:parameter} we present results investigating the effect of introducing additional parameters in QAOA.

For all numerical experiments presented in this Section, we use \texttt{QAOAKit} \cite{shaydulin2021qaoakit} to optimize the QAOA parameters. We use L\_BFGS\_B \cite{liu1989limited} as the optimizer. The algorithm is a quasi-Newton method to minimize the value of a differentiable scalar function $f$. The derivatives of $f$ are used to identify the direction of the steepest descent and also to estimate the Hessian of $f$. We use the implementation available in the Scipy~\cite{scipy} package with the default setting, we set the maximum number of iterations to 20000 to guarantee convergence of the optimizer. Following the methodology in \cite{khairy2019learning,shaydulin2022parameter}, we run the local optimizer from 10 initial points sampled from a generative model trained using Kernel Density Estimation on optimized parameters for MaxCut on all non-isomorphic 9-node graphs, in addition to 30 initial points sampled uniformly randomly from $[-2\pi, 2\pi]$. We make the implementation available on GitHub at \href{https://github.com/JoeyXLiu/qaoa-sparse}{https://github.com/JoeyXLiu/qaoa-sparse}.

\subsection{Motivation}\label{sec:motivation}
In standard QAOA for MaxCut, the phase operator is formulated directly using the graph, namely, for each edge $(i, j) \in E$, there is an operator 
\[
e^{-i\gamma\sigma_i^z\sigma_j^z}
\]
that acts on qubits $i$ and $j$ in the phase operator. Such phase operator may require a large number of gates that potentially do not match the hardware connectivity.

\begin{figure*}[htbp]
\centering
\includegraphics[width=\linewidth]{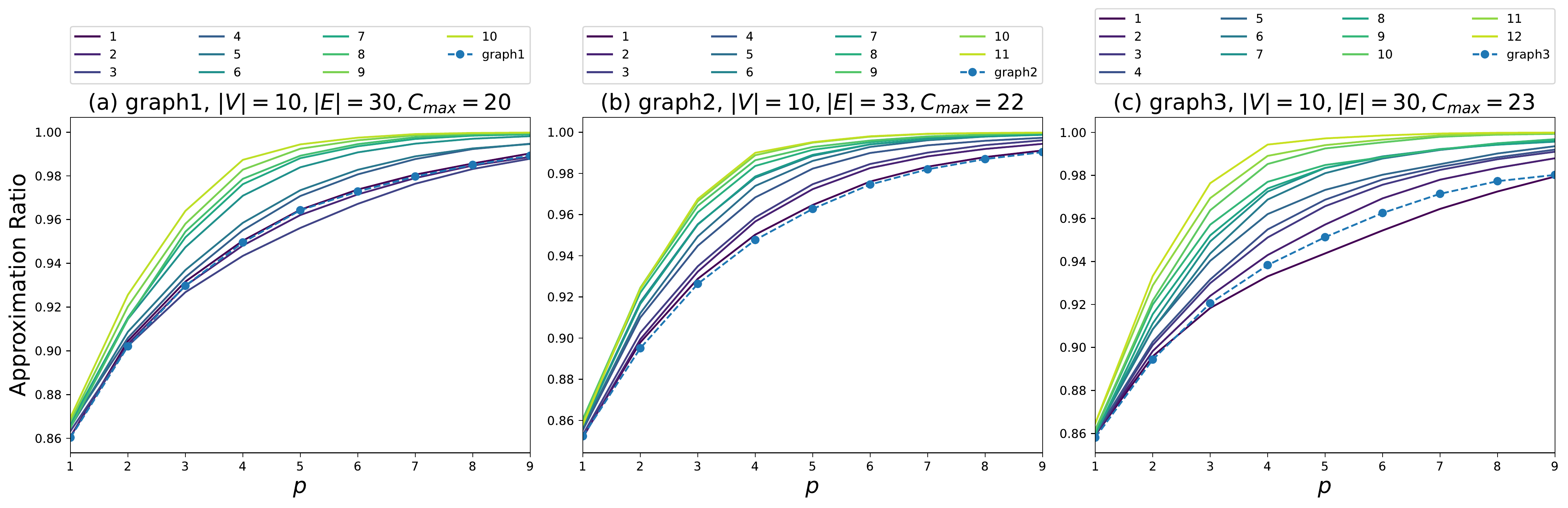}
\caption{Comparison of Sparse QAOA and standard QAOA. Edges that are not in the cut are deleted gradually. The index of the sparsified graph indicates how many edges we delete. The horizontal axis corresponds to the QAOA circuit depth. The vertical axis shows an approximation ratio  $\rho$.}\label{fig:delete}
\end{figure*}

We propose to combine graph sparsification and QAOA, which starts with a graph sparsification process that removes some edges from the graph, and then uses the sparsified graph to formulate the phase operator. The more edges we remove, the fewer gates we have in the phase operator. As an motivation example, we start with the following numerical experiments. We generate three 10-node random graphs, with 30, 33, and 35 edges respectively. Since the the size of the graphs is relatively small, we can find the optimal solutions. Given a solution of MaxCut, we can classify the edges into two sets, the edges that are in the cut, i.e., the edges that span two parts, and the edges that are not in the cut, i.e., all other edges. We gradually delete the edges that are not in the cut. The results are presented in Fig.~\ref{fig:delete}. The index of the sparsified graph indicates how many edges we delete. \emph{Remarkably, most of the sparse QAOA outperform standard QAOA, obtaining higher approximation ratio compared with standard QAOA. Moreover, the QAOA equipped with the phase operator corresponding to the graph with the least number of edges (which requires the least gates to implement) obtains the best result among all.} 

\subsection{Sparse QAOA and Random Sparse QAOA}\label{sec:sparseqaoamaxcut}

The observation in Section \ref{sec:motivation} inspires us to design an algorithm to sparsify the QAOA circuit by removing the edges that are most likely not in the cut. For example, we can start from an initial solution obtained by fast  classical algorithms, and then sparsify the graph based on this initial solution. We describe our algorithm in the next subsection. We outline this approach in Algorithm \ref{alg:sparse_qaoa}.

\begin{algorithm}[H]
\begin{algorithmic}[1] 
\caption{Sparse QAOA}
\label{alg:sparse_qaoa}
\STATE Generate an initial Max-Cut solution using non-quantum fast heuristic
\STATE Sparsify the graph by removing the edges that are not in the cut
\STATE Use the sparsified graph as the Hamiltonian in the phase operator of QAOA circuit
\STATE Optimize parameters to maximize $\bra{\gamma, \beta} \mathcal{H} \ket{\gamma, \beta}$ until convergence, where $\mathcal{H}$ is the Hamiltonian for the original graph.
\end{algorithmic}
\end{algorithm}

\begin{figure*}[htbp]
\centering
\includegraphics[width=\linewidth]{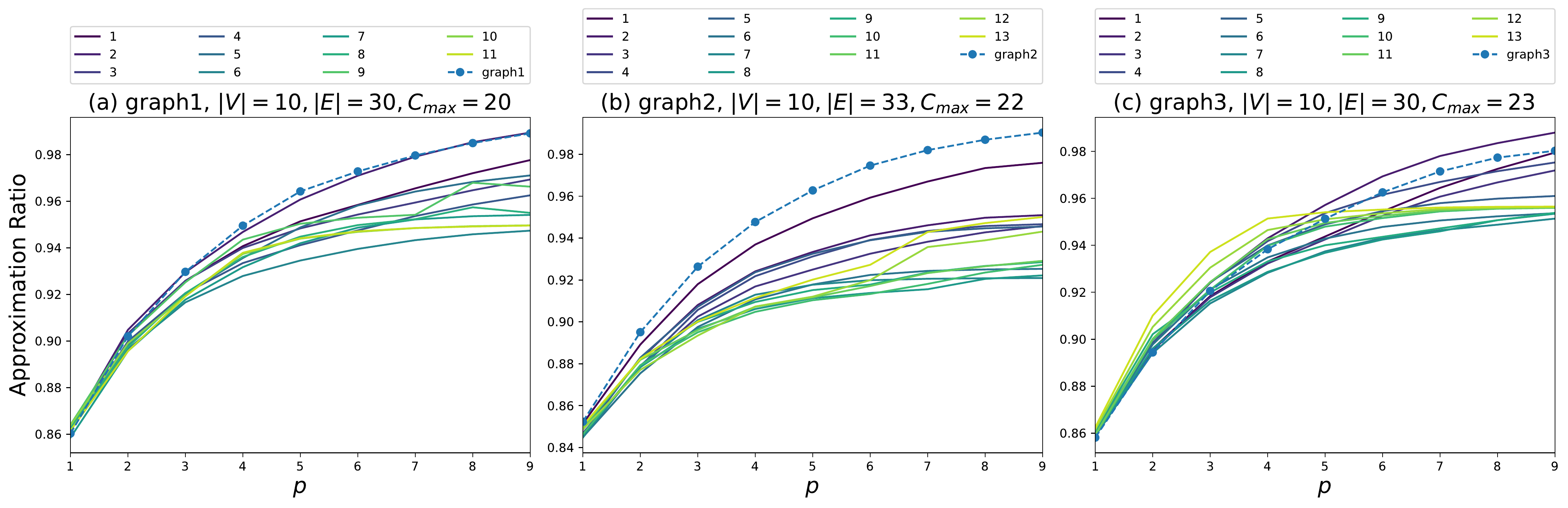}
\caption{Sparse QAOA initialized with Goemans-Williamson algorithm solution and standard QAOA. Edges that are not in the cut in the Goemans-Williamson algorithm solution are deleted gradually. The index of the sparsified graph indicates how many edges we delete.}\label{fig:goemans}
\end{figure*}

To generate the initial solution in step 1 of Algorithm \ref{alg:sparse_qaoa}, we can use any classical efficient heuristic or approximation algorithm. One such possible choice could be the best-known approximation algorithm for MaxCut, the Goemans-Williamson algorithm \cite{goemans1995improved}. This algorithm achieves an approximation ratio of 0.878. We test the same three random graphs again when the Goemans-Williamson algorithm gives a solution that is not optimal. Again, we gradually delete the edges that are not in the cut in the initial solution. The index of the sparsified graph indicates how many edges we delete. We present the results in Fig.~\ref{fig:goemans}. We now observe that only some of the sparse QAOA have comparable performance compared to the standard QAOA. On one hand, they require fewer gate operations since the edges that are not in the cut are removed from the phase operator. However, we observe that when the initial solution is not optimal, we inevitably remove edges that are in the optimal cut. When such edges are deleted, the results of sparse QAOA deteriorate.

To improve the algorithm, instead of removing all edges that are not in the cut as in Algorithm \ref{alg:sparse_qaoa}, we randomly select a subset of these edges to remove. We outline this algorithm in Algorithm \ref{alg:random_sparse_qaoa}.

\begin{algorithm}[H]
\begin{algorithmic}[1] 
\caption{Random Sparse QAOA}
\label{alg:random_sparse_qaoa}
\STATE Generate an initial Max-Cut solution
\STATE Randomly select edges that are not in the cut with probability $p_e$.
\STATE Sparsify the graph by removing the selected edges.
\STATE Use the sparsified graph as the Hamiltonian in the phase operator of QAOA circuit.
\STATE Optimize parameters to maximize $\bra{\gamma, \beta} \mathcal{H} \ket{\gamma, \beta}$ until convergence.
\end{algorithmic}
\end{algorithm}

\begin{figure*}[htbp]
\centering
\includegraphics[width=\linewidth]{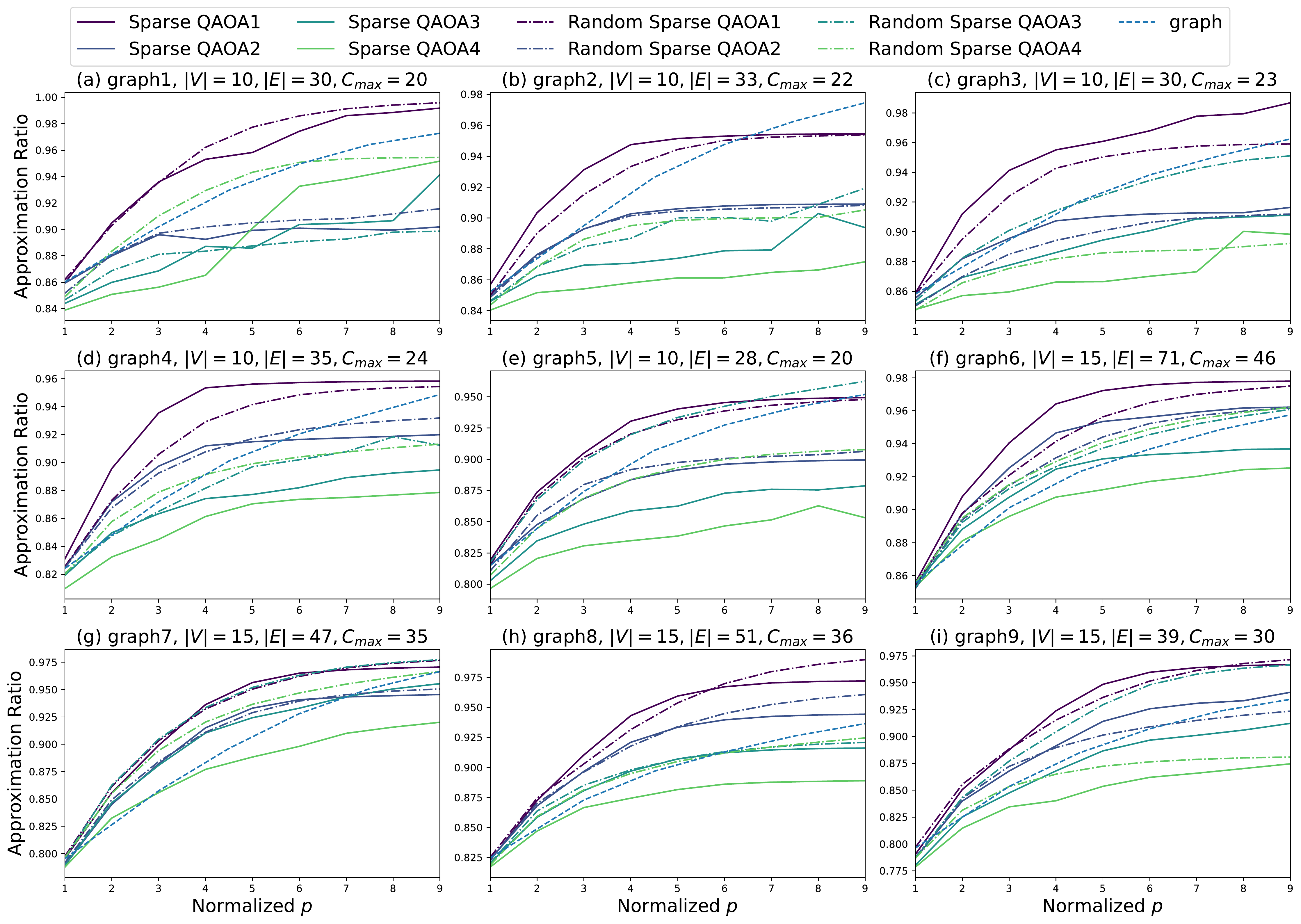}
\caption{Comparison of Sparse QAOA, random Sparse QAOA and standard QAOA. The initial solutions used in Sparse QAOA and random Sparse QAOA are manually picked and the index indicates how far away the initial solution is away from the optimal solution. The $p$ value of the standard QAOA is scaled with respect to the gate counts for easier comparison. }\label{fig:sparse_random}
\end{figure*}

We now compare the performance of Sparse QAOA and Random Sparse QAOA. To investigate the impact of the initial solution, we manually choose the initial solution. For example, graph1 has 30 edges and the optimal cut contains 20 edges, we manually pick an initial solution that has cut 19, 18, 17, and 16 edges respectively. The index in the plots indicates how far away the initial solution is from the optimal solution of the original graph. The results are presented in Fig.~\ref{fig:sparse_random} (a) - (c). Note that in the plots with graph sparsification, we remove approximately 1/3 or more of the edges, so we scale and normalize the $p$ value of the standard QAOA based on the gate count in the phase operator for easier comparison. Namely, the gate count in the phase operator on the original graph for $p = 2$ is roughly equivalent to the gate count of QAOA on the sparsified graph for $p = 3$, similarly, the gate count in the phase operator on the original graph for $p = 4$ is roughly equivalent to the gate count of QAOA on the sparsified graph for $p = 6$, the gate count in the phase operator on the original graph for $p = 6$ is roughly equivalent to the gate count of QAOA on the sparsified graph for $p = 9$. When we start from a good initial solution, i.e., sparse QAOA1 and random sparse QAOA1 can have comparable performance or even better performance compared to standard QAOA while using fewer gates in the phase operator. When the solution is far away from the optimal solution, sparse QAOA is having poor performance, and random sparse QAOA can improve the results of sparse QAOA. We test more graphs of sizes 10 and 15, and observe similar results. The results are shown in Fig.~\ref{fig:sparse_random} (d) - (i).

The main advantage of sparse QAOA and random sparse QAOA is that they require fewer gates than standard QAOA. Since for each edge $(i, j) \in E$ removed in the sparsification, we drop a $e^{-i\gamma\sigma_i^z\sigma_j^z}$ operator that acts on qubit $i$ and $j$ in the phase operator. The more edges we remove, the fewer gates we have in the phase operator. In the absence of error correction, reduced gate count implies higher fidelity of the final state.

\subsection{Performance of Sparse QAOA}\label{sec:sparsification}
\begin{figure*}[htbp]
\centering
\includegraphics[width=\linewidth]{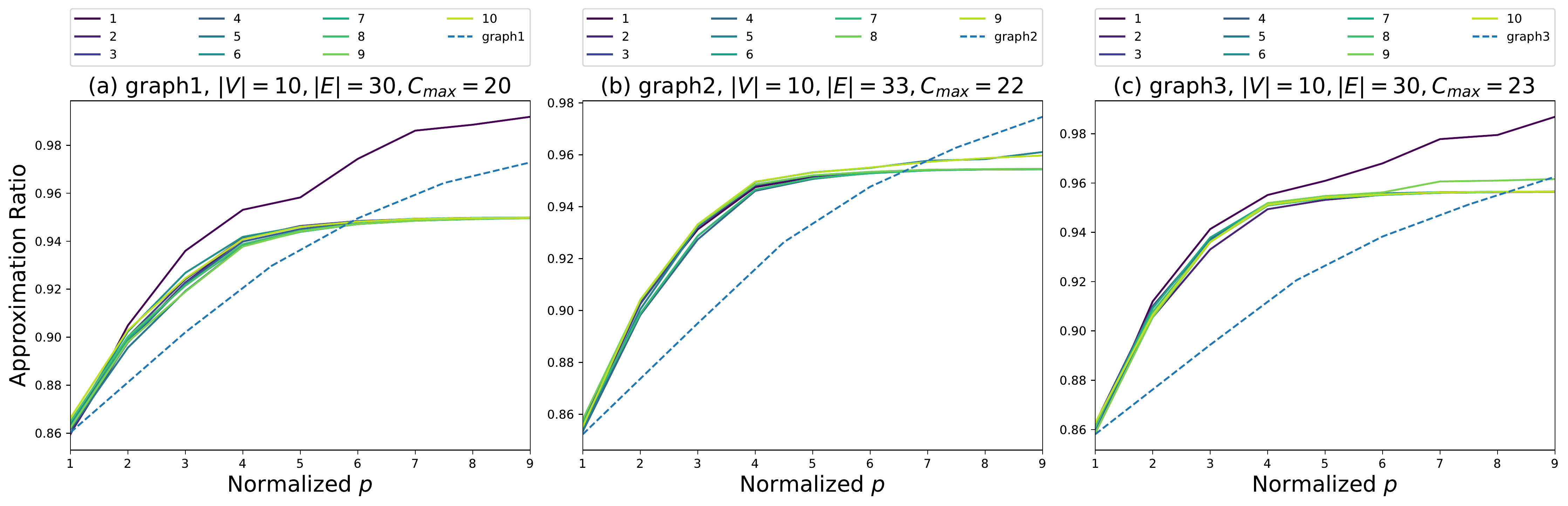}
\caption{Sparse QAOA initialized with different solutions that gives the same cut. These solutions give objective value $C_{\max} - 1$ and edges that are not in the cut are deleted. The $p$ value of the standard QAOA is scaled with respect to the gate count for easier comparison.}\label{fig:sparse_suboptimal}
\end{figure*}

\paragraph{Varying initial solution}
We investigate the performance of sparse QAOA given different initial solutions. For each graph, there are multiple solutions that give the same cut. Even though they have the same objective function value, the edges that are not in the cut can differ dramatically between different solutions. We sparsify the graph based on these different solutions. The results are presented in Fig.~\ref{fig:sparse_suboptimal}. We observe that initial solutions that give the same cut generally have similar performance.

\begin{figure*}[htbp]
\centering
\includegraphics[width=\linewidth]{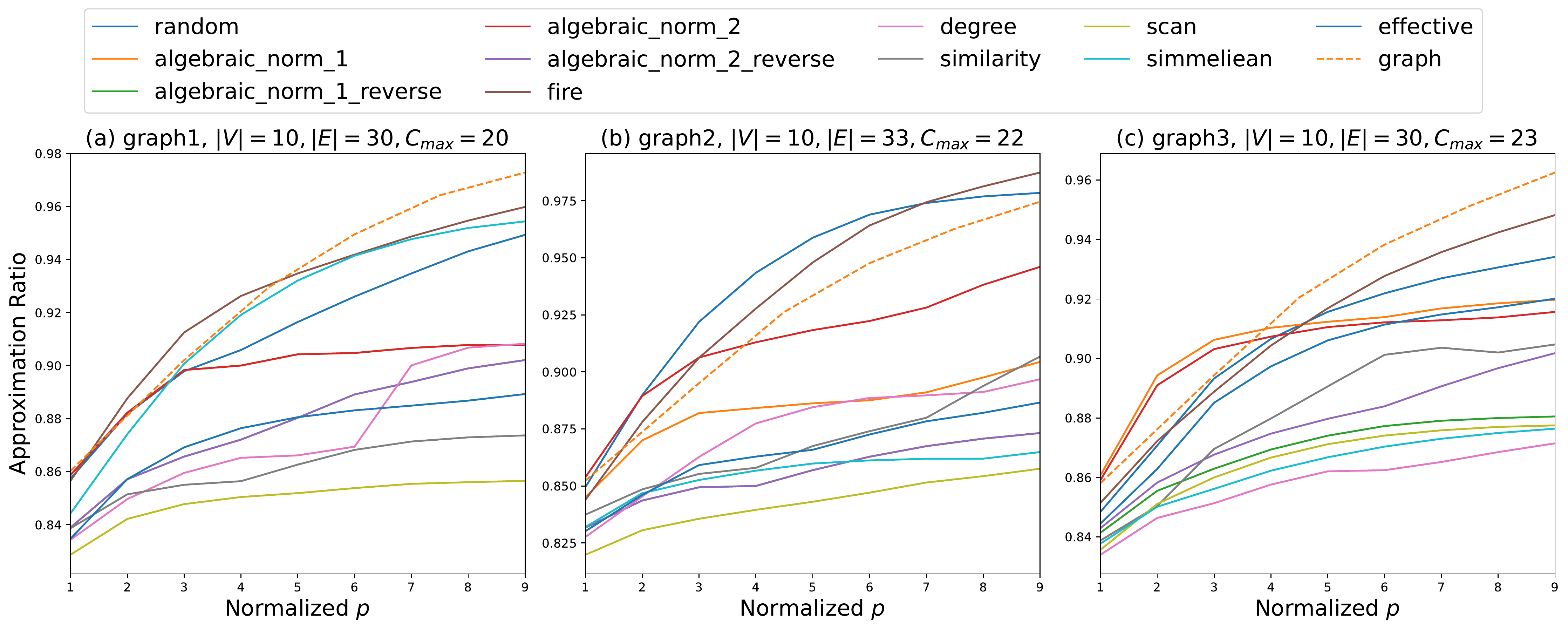}
\caption{Comparison of Sparse QAOA with different sparsification methods and standard QAOA. The graphs are sparsified with target ratio 66\%. The $p$ value of the standard QAOA is scaled with respect to the gate count for easier comparison.}\label{fig:sparse}
\end{figure*}

\paragraph{Varying graph sparsification methods}
We now compare the performance of QAOA with different graph sparsification methods reviewed in Section \ref{sec:sparsification}. We first sparsify the graph with different graph sparsification approaches. For random edge, algebraic distance, edge forest fire, local degree, local similarity, SCAN structural similarity and Simmelian backbones, we use the implementation provided in \texttt{Networkit}~\cite{staudt2016networkit,hamann2016structure}. For effective resistance, we use the \texttt{gSparse}~\cite{gsparse} implementation. The target ratio is 66\%, i.e., the sparsified graph will preserve 66\% of the edges. Then we use the sparsified Hamiltonian in the QAOA circuit. The results are presented in Fig.~\ref{fig:sparse}. Note that here in the plots, we again scale and normalize the $p$ value of the standard QAOA based on the gate count in the phase operator for easier comparison. We observe that, the sparsified QAOA with $p=3$ can sometimes reach a better approximation ratio compared to the standard QAOA with $p = 2$. These numerical results suggest that the removal of some edges from the phase operator, does not affect the performance of QAOA significantly, and in some cases can achieve comparable performance with the standard QAOA while using fewer gates. However, there is not a clear winner when we compare all sparsification methods.

\subsection{Energy levels alignment and QAOA performance}\label{sec:en_alignment}

We now investigate the relationship between the QAOA approximation ratio and how well the sparsified graph approximates the original one. In the adiabatic limit, as long as the Hamiltonian encoding the cut of the sparsified graph has the same ground state as the original one, the Sparse QAOA will still obtain the optimal cut. Since we consider QAOA with small depth $p$, this limit does not apply directly. However, we still observe that if the optimal solutions of the sparsified graph correspond to the optimal solutions of the original graph, the approximation ratio obtained by Sparse QAOA is on average better than that of QAOA with original phase operator.

Moreover, we observe that better alignment in energy levels between the Hamiltonians encoding the original and sparsified graphs leads to better Sparse QAOA performance. We compute this alignment in the following way. We sort the binary strings encoding the solutions in order of decreasing cut value for both the original and the sparsified graph, with $1$st energy level corresponding to the optimal cut. Denote the set of all binary strings corresponding to $k$th largest cut for graph $G$ as $S(k,G)$. Then the $k$th energy level of $G$ and $G'$ are aligned if either $S(k,G)\subseteq S(k,G')$ or  $S(k,G')\subseteq S(k,G)$. The number of aligned energy levels is equal to the largest $k$ such that $k$th energy levels are aligned. 

\begin{figure}
    \centering
    \includegraphics[width=0.8\linewidth]{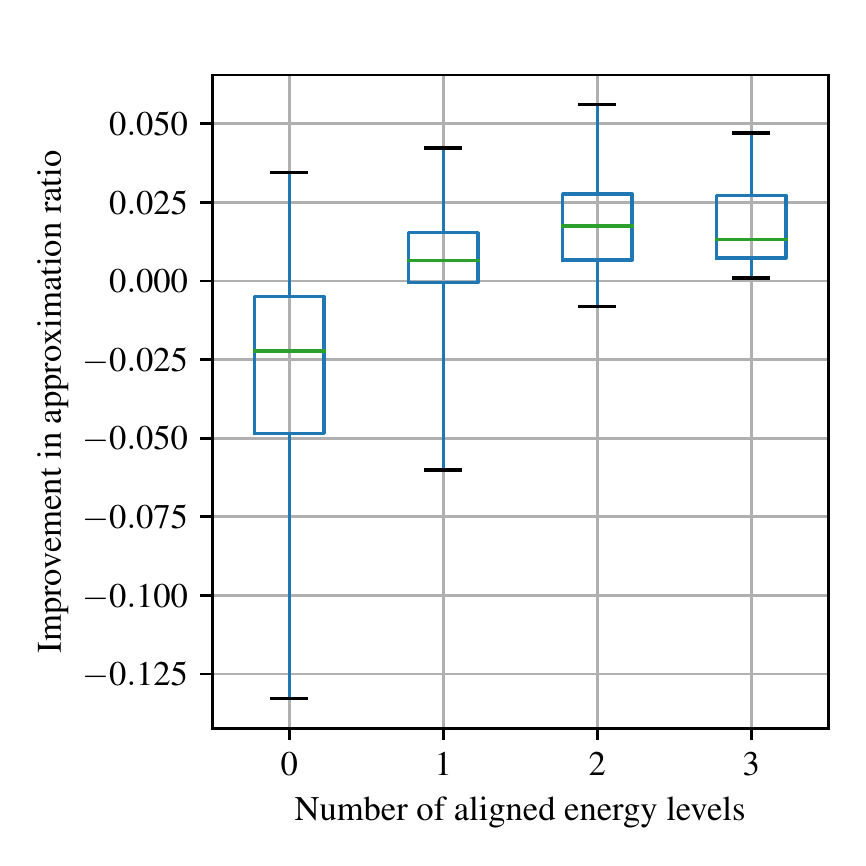}
    \caption{The difference in approximation ratio between QAOA with the sparsified phase operator and original one as a function of the number of aligned energy levels. Positive values indicate that Sparse QAOA gives better approximation ratio.}
    \label{fig:energy_lvl_alignment}
\end{figure}

The relationship between the number of aligned energy levels and the difference in QAOA approximation ratio is presented in Fig.~\ref{fig:energy_lvl_alignment}.  As the number of aligned energy levels increases, the improvement in QAOA approximation ratio from using the sparsified phase operator increases. Note that if the number of aligned energy levels is not zero, the optimal solutions of the sparsified graph are also optimal for the original graph (i.e. the ground state is aligned). Interestingly, we observe that aligning more than one low energy state gives better performance, suggesting that the entire low-energy subspace plays a role.

\subsection{Additional Parameters in Phase Operator}\label{sec:parameter}
\begin{figure*}[htbp]
\centering
\includegraphics[width=\linewidth]{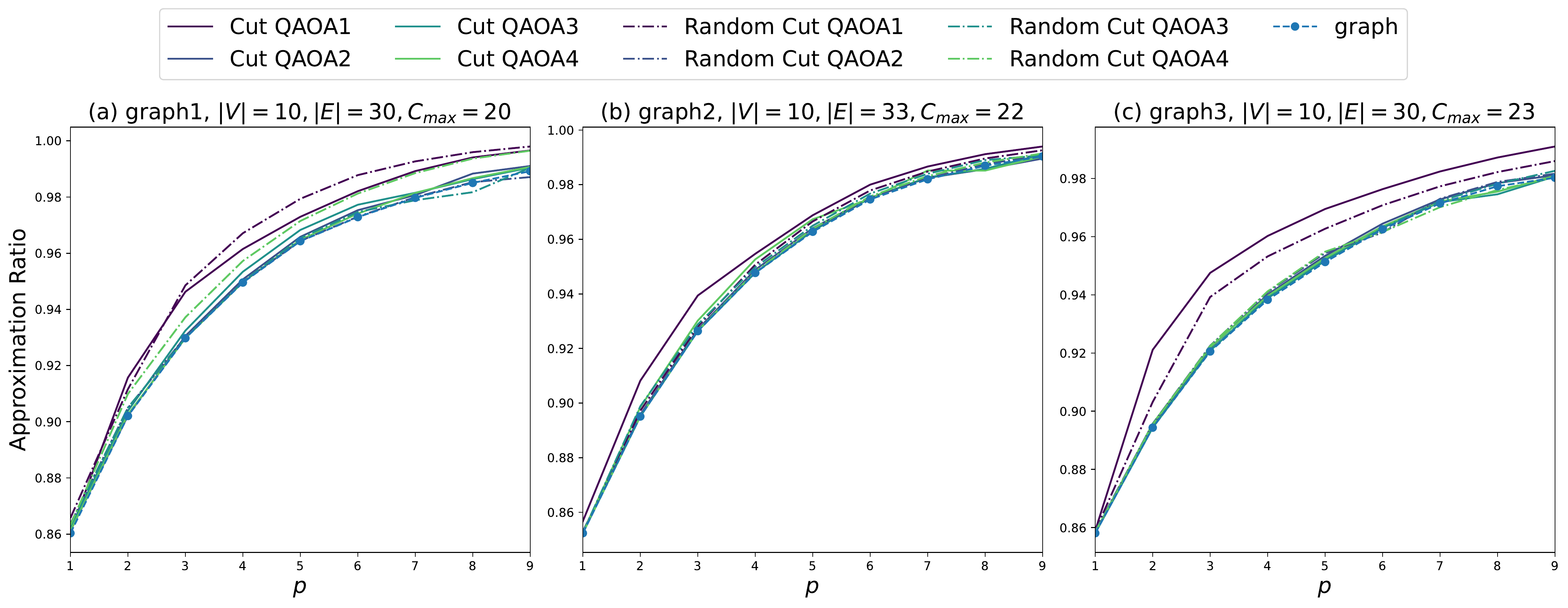}
\caption{Comparison of Cut QAOA, random Cut QAOA and standard QAOA. The initial solutions used in Cut QAOA and random Cut QAOA are manually picked and the index indicates how far away the initial solution is away from the optimal solution. }\label{fig:cut_qaoa}
\end{figure*}

Previous works claim that introducing additional classical parameters can usually improve the performance of QAOA. For example in \cite{herrman2021multi}, the authors introduce additional parameters for each edge in the phase operator and those for each node in the mixing operator, their ma-QAOA gives better approximation ratio than QAOA. Therefore, for the next set of experiments, we consider using the same quantum resources as standard QAOA, but introducing additional parameters to the phase operator. Namely, given an initial solution, we partition the edges into two sets, and assign $\gamma_{p1}$ to the edges that are in the cut, and $\gamma_{p2}$ to the edges that are not in the cut. We name this approach Cut QAOA, outlined in Algorithm \ref{alg:cut_qaoa}. The standard QAOA will be a special case of Cut QAOA when $\gamma_{p1} = \gamma_{p2}$, therefore, the performance of Cut QAOA should be at least good as standard QAOA. We also test adding a randomized procedure similar to Random Sparse QAOA and name this approach Random Cut QAOA. The algorithm is outlined in Algorithm \ref{alg:random_cut_qaoa}. The index in the plots again indicate how far away the initial solution is from the optimal solution. The results are presented in Fig.~\ref{fig:cut_qaoa}. Indeed, Cut QAOA and Random Cut QAOA outperform standard QAOA.

\begin{algorithm}[H]
\begin{algorithmic}[1] 
\caption{Cut QAOA}
\label{alg:cut_qaoa}
\STATE Generate an initial Max-Cut solution.
\STATE Assign $\gamma_{p1}$ to the edges that are in the cut, and $\gamma_{p2}$ to the edges that are not in the cut to formulate the Hamiltonian in the phase operator of QAOA circuit.
\STATE Optimize parameters to maximize $\bra{\gamma, \beta} \mathcal{H} \ket{\gamma, \beta}$ until convergence.
\end{algorithmic}
\end{algorithm}

\begin{algorithm}[H]
\begin{algorithmic}[1] 
\caption{Random Cut QAOA}
\label{alg:random_cut_qaoa}
\STATE Generate an initial Max-Cut solution.
\STATE Randomly select edges that are not in the cut with probability $p_e$, we denote this subset of edges as $S$.
\STATE Assign $\gamma_{p1}$ to the edges $e\in S$, and $\gamma_{p2}$ to the rest of edges $e\not\in S$ to formulate the Hamiltonian in the phase operator of QAOA circuit.
\STATE Optimize parameters to maximize $\bra{\gamma, \beta} \mathcal{H} \ket{\gamma, \beta}$ until convergence.
\end{algorithmic}
\end{algorithm}

\section{Discussion}\label{sec:discussion}
 QAOA is a promising candidate to demonstrate quantum advantage on near-term devices.  However, the implementation of QAOA on hardware is significantly affected by the connectivity of the hardware and gates fidelity. In this work, we propose to sparsify the graph based on an initial solution of MaxCut, and use the sparsified graph to formulate the phase operator in the QAOA circuit. As such, the number of gates used in the QAOA circuit will be reduced. Furthermore, we show that sparse QAOA can have comparable performance and can even sometimes outperform the standard QAOA if the ground state is preserved, especially when $p$ is small. The initial solution can be produced by any fast classical algorithm. One good choice is the Goemans-Williamson algorithm. When the initial solution is far away from the optimal solution, we further introduce a randomized procedure to improve sparse QAOA. Numerical results show that random sparse QAOA can again have a comparable performance and sometimes outperform the standard QAOA. We also compare different graph sparsification methods, although some of them occasionally achieve comparable performance, there is no clear winner. Investigation on other efficient and robust graph sparsification methods that lead to improving QAOA is an interesting future research direction that can potentially lead to superior and accelerated QAOA.
 
 We observe that the energy spectrum of the Hamiltonian representing the sparsified graph needs to align well with that of the original graph to preserve or improve the QAOA approximation ratio. This provides evidence against the idea that any Hamiltonian that does not commute with the mixer can be used in QAOA, which has been suggested to be a sufficient condition previously~\cite{hadfield2017quantum}. More research is needed to identify the properties of the phase operator that lead to improved QAOA performance.
 
 An independent paper studying the effects of removing some clauses from the problem Hamiltonian has appeared on arXiv~\cite{wang2022dropout} during the final stages of preparation of this manuscript. Similar observations have been found for Not-all-equal 3-SAT problem.

\section*{Acknowledgments}

This research was supported in part through the use of DARWIN computing system: DARWIN – A Resource for Computational and Data-intensive Research at the University of Delaware and in the Delaware Region, Rudolf Eigenmann, Benjamin E. Bagozzi, Arthi Jayaraman, William Totten, and Cathy H. Wu, University of Delaware, 2021, URL: https://udspace.udel.edu/handle/19716/29071.
We gratefully acknowledge the computing resources provided on Bebop, a high-performance computing cluster operated by the Laboratory Computing Resource Center at Argonne National Laboratory. This work was supported in part with funding from the ONISQ program of the Defense Advanced Research Projects Agency (DARPA). The views, opinions and/or findings expressed are those of the authors and should not be interpreted as representing the official views or policies of the Department of Defense or the U.S. Government.

\bibliographystyle{IEEEtran}
\bibliography{ref}

\vfill
\framebox{\parbox{.90\linewidth}{\scriptsize The submitted manuscript has been created by
UChicago Argonne, LLC, Operator of Argonne National Laboratory (``Argonne'').
Argonne, a U.S.\ Department of Energy Office of Science laboratory, is operated
under Contract No.\ DE-AC02-06CH11357.  The U.S.\ Government retains for itself,
and others acting on its behalf, a paid-up nonexclusive, irrevocable worldwide
license in said article to reproduce, prepare derivative works, distribute
copies to the public, and perform publicly and display publicly, by or on
behalf of the Government.  The Department of Energy will provide public access
to these results of federally sponsored research in accordance with the DOE
Public Access Plan \url{http://energy.gov/downloads/doe-public-access-plan}.}}
\clearpage
\end{document}